\documentclass[10pt,pra,twocolumn,superscriptaddress,showpacs,floatfix]{revtex4}

\usepackage{graphicx}
\usepackage{amssymb}
\usepackage{times}
\usepackage{amsmath}
\usepackage{amsthm}

\input epsf.tex
\usepackage{graphicx}
\usepackage{amsthm}

\begin{document}

\title{Quantum Hacking on Continuous-Variable Quantum Key Distribution System using a Wavelength Attack}

\author{Jing-Zheng Huang}\affiliation
 {Key Laboratory of Quantum Information, University of Science and Technology of China, Hefei, 230026, China}

\author{Christian Weedbrook}\affiliation{Center for Quantum Information and Quantum Control,
Department of Electrical and Computer Engineering and Department of Physics, University of Toronto, Toronto, M5S 3G4, Canada}

\author{Zhen-Qiang Yin\footnote{yinzheqi@mail.ustc.edu.cn}}\affiliation
 {Key Laboratory of Quantum Information, University of Science and Technology of China, Hefei, 230026, China}

\author{Shuang Wang\footnote{wshuang@ustc.edu.cn}}\affiliation
 {Key Laboratory of Quantum Information, University of Science and Technology of China, Hefei, 230026, China}

\author{Hong-Wei Li}\affiliation
 {Key Laboratory of Quantum Information, University of Science and Technology of China, Hefei, 230026, China}

\author{Wei Chen}\affiliation
 {Key Laboratory of Quantum Information, University of Science and Technology of China, Hefei, 230026,  China}

\author{Guang-Can Guo}\affiliation
 {Key Laboratory of Quantum Information, University of Science and Technology of China, Hefei, 230026,  China}

\author{Zheng-Fu Han}\affiliation
 {Key Laboratory of Quantum Information, University of Science and Technology of China, Hefei, 230026, China}

\date{\today}

\begin{abstract}

The security proofs of continuous-variable quantum key distribution are based on the assumptions that the eavesdropper can neither act on the local oscillator nor control Bob's beam splitter. These assumptions may be invalid in practice due to potential imperfections in the implementations of such protocols. In this paper, we consider the problem of transmitting the local oscillator in a public channel and propose a wavelength attack which can allow the eavesdropper to control the intensity transmission of Bob's beam splitter by switching the wavelength of the input light. Specifically we target continuous-variable quantum key distribution systems that use the heterodyne detection protocol using either direct or reverse reconciliation. Our attack is proved to be feasible and renders all of the final key shared between the legitimate parties insecure, even if they have monitored the intensity of the local oscillator. To prevent our attack on commercial systems, a simple wavelength filter should be randomly added before performing the monitoring detection.

\end{abstract}

\maketitle

\section{Introduction}\label{Introduction}

Quantum key distribution~(QKD) enables two distant partners, Alice and Bob, to share common secret keys in the presence of an eavesdropper, Eve~\cite{Sca09,qkd}. In theory, the unconditional security of QKD protocol is guaranteed based on the laws of physics, in particular the no-cloning theorem. But in practice, the key components of practical QKD systems have imperfections that do not fulfill the assumptions of ideal devices in theoretical security proofs. In discrete-variable QKD, the imperfect devices such as single photon detectors, phase modulators, Faraday mirrors and fiber beam splitters, open security loopholes to Eve and lead to various types of attacks~\cite{Qi2007,Zhao2008,Lydersen2010,Gerhardt2011,Wiechers2011,Lars,Xu,Weier,Jain,WaveAtt}.

Continuous-variable~(CV) QKD~\cite{Weedbrook2011} has developed immensely over the past decade~\cite{Jouguet2012} to the point that there are companies selling commercially available systems~\cite{quintessencelabs,securenet}. Even so, CV-QKD is potentially vulnerable to such idealization-to-practical problems that plague its discrete variable counterpart. In the CV-QKD protocols, Alice encodes the key information onto the quadratures,  $\hat{X}$ and $\hat{P}$, on a bunch of coherent states and sends them onto Bob. Bob measures one or both quadratures by performing homodyne~\cite{cv} or heterodyne detection~\cite{cv-het} on the signal with a relatively strong local oscillator~(LO). Finally, they perform direct or reverse reconciliation and privacy amplification process to distill a common secret key~\cite{Weedbrook2011,Sca09}. In practice, it is extremely difficult for Bob to generate the LO with the same initial polarization and phase to Alice's signal. Therefore, Alice prepares both the signal and LO, and send them to Bob in the same optical fiber channel at the same time to avoid the large drifts of the relative polarization and phase~\cite{cv-exp}. However, this implementation leaves a security loophole open for Eve.

In Ref.~\cite{testlo}, the authors proposed an equal-amplitude attack. To perform this attack, Eve first intercepts the signal and LO, and measures both of the quadratures by performing heterodyne detection on them. According to her measurement results, she reproduces two weak squeezed states which have the same intensity level to the signal, and sends them onto Bob. Bob treats these two fake states as signal and LO, and performs detections on them as usual. But now the detection is neither homodyne nor heterodyne detection, therefore Eve is able to make the extra noise of Bob's measurement much lower than the shot noise level. As a result, the total deviation between Bob's measurement and Alice's preparation is lower than the tolerable threshold derived from the theoretical security proofs~\cite{cv-sec,het-sec}. Hence Alice and Bob can not discover the presence of Eve.

In order to prevent this attack without modifying the original measurement setup, Bob needs to monitor the total intensity or the LO intensity~\cite{testlo}. We note that in this attack, Eve is assumed to be unable to control the beam splitters of Bob. But in one of our recent studies~\cite{WaveAtt}, we found that it is possible for Eve to control the outputs of fiber beam splitters by utilizing its wavelength dependent property~\cite{bs-0,bs-1,bs-2}. Importantly, such wavelength dependent properties can be found in commercial CV-QKD systems~\cite{quintessencelabs,securenet}. Making use of this loophole, we propose a new wavelength attack on a practical CV-QKD system using heterodyne detection protocol~\cite{cv-het}. By using this attack Eve can in principle achieve all of the secret key without being discovered, even if Bob has monitored the total intensity or the LO intensity. Such an attack has practical and commercial consequences.

In the security analysis of CV-QKD protocols with direct (reverse) reconciliation, $V_{A|B}$($V_{B|A}$), Alice (Bob)'s conditional variance of Bob (Alice), has the similar status as the quantum bit error rate~(QBER) in the discrete-variable QKD protocols. To show that the hidden Eve would not be discovered in our attack, our method is proving that the upper bound of $V_{A|B}$~($V_{B|A}$) under our wavelength attack is always lower than the maximum value allowed by the secret key rate formula~\cite{cv-het,het-sec}.

This paper is organized as follows. In Section~\ref{Preliminary}, we first review the heterodyne protocol and the wavelength-dependent property of certain fiber beam splitters, then we propose a wavelength attack scheme on an all-fiber CV-QKD system using heterodyne protocol in Section~\ref{Wavelength Attack}. We prove the feasibility of this wavelength attack in Section~\ref{Feasibility Analysis}, and finally conclude in Section~\ref{conclusion}.

\section{Preliminary}\label{Preliminary}

\subsection{Heterodyne detection protocol}\label{HDP}

\begin{figure}[!h]\center
\resizebox{9cm}{!}{
\includegraphics{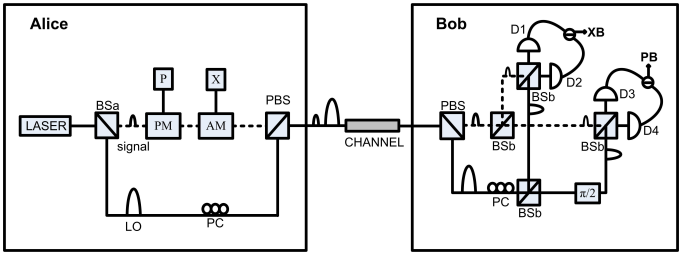}}
\caption{The schematic diagram of heterodyne detection protocol. BSa: $1/99$ beam splitter; BSb: 50/50 beam splitter; PM: phase modulator; AM: amplitude modulator; PBS: polarization beam splitter; PC: polarization controller. Alice generates coherent light pulses by a $1550$~nm laser diode, then separates them into a weak signal and a strong LO by a $1/99$ beam splitter. The signal is then modulated randomly following the centered Gaussian distribution in both quadratures, by using phase and amplitude modulators. The signal and LO are separated in time and modulated into orthogonal polarizations by the PBS before begin inserted into the channel.}\label{figure schematic 1}
\end{figure}

In the heterodyne protocol~\cite{cv-het}, Alice first prepares a displaced vacuum state that will be sent to Bob. This is realized by choosing two real numbers $X_A$ and $P_A$ from a Gaussian distribution of variance $V_A$ and zero mean. The whole ensemble of coherent states Alice will send to Bob is given by the thermal state with variance $V = V_A +1$. Bob receives this coherent state and simultaneously measures both the amplitude and phase quadratures of the state using heterodyne detection. After repeating this process many times, they finally extract a binary secret key by using either direct reconciliation~\cite{Grosshans2002} or reverse reconciliation algorithm~\cite{cv-het}. A typical CV-QKD system using heterodyne protocol can be realized by the schematic shown in Fig.~\ref{figure schematic 1}. In this scheme, time and polarization multiplexing are used so that the signal and LO can be transmitted in the same channel without interfering. To avoid the equal-amplitude attack~\cite{testlo}, Bob uses a 10:90 beam splitter(not depicted in the figure) before the polarization beam splitter to monitor the LO intensity\cite{Jouguet2012}.

To perform the heterodyne detection, Bob uses the photo-detector to convert the photons into a photocurrent $\hat{i}$. Here $\hat{i}$ and the photon number $\hat{n}$ are related by $\hat{i} = q\hat{n} = q\hat{a}^{\dag}\hat{a}$, where $\hat{a}$ and $\hat{a}^{\dag}$ are the annihilation and creation operators of the light state, and q is a suitable constant~\cite{cv-rev}. The extra quantum noise $\delta\hat{\alpha}_v$ is unavoidable in Bob's measurement results when he uses heterodyne detection due to the unused port of the $50:50$ beam splitter. To show this, let us first describe the signal and LO by operators $\hat{\alpha}_s$ and $\hat{\alpha}_{LO}$, respectively. These operators can be broken up into two contributions~\cite{book}: the mean values of the amplitude $\alpha$ as well as the quantum noise fluctuations $\delta \alpha$. The operators can be written as
\begin{equation}
\begin{array}{lll}
\hat{\alpha}_s = \alpha_s + \delta\hat{\alpha}_s,\\
\hat{\alpha}_{LO} = \alpha_{LO} + \delta\hat{\alpha}_{LO}.
\end{array}
\end{equation}
where $\alpha_s$ and $\alpha_{LO}$ are complex numbers and we assume that the amplitude of the LO is much larger than the signal, i.e., $|\alpha_{LO}| \gg |\alpha_s|$, and $\delta\hat{\alpha}_s$ and $\delta\hat{\alpha}_{LO}$ are the fluctuations of the signal and LO, respectively.

The photocurrents read by the four photo-detectors can be written as follows
\begin{equation}
\begin{array}{lll}
\hat{i}_1 &= q(\alpha_{LO}^* + \delta\hat{\alpha}_{LO}^{'\dag} + \alpha_s^* + \delta\hat{\alpha}_s^{'\dag})\times\\&(\alpha_{LO} + \delta\hat{\alpha}'_{LO} + \alpha_s + \delta\hat{\alpha}'_s)/4,\\
\hat{i}_2 &= q(\alpha_{LO}^* + \delta\hat{\alpha}_{LO}^{'\dag} - \alpha_s^* - \delta\hat{\alpha}_s^{'\dag})\times\\&(\alpha_{LO} + \delta\hat{\alpha}'_{LO} - \alpha_s - \delta\hat{\alpha}'_s)/4,\\
\hat{i}_3 &= q[e^{-i\frac{\pi}{2}}(\alpha_{LO}^* + \delta\hat{\alpha}_{LO}^{'\dag}) + \alpha_s^* + \delta\hat{\alpha}_s^{'\dag}]\times\\&[e^{i\frac{\pi}{2}}(\alpha_{LO} + \delta\hat{\alpha}'_{LO}) + \alpha_s + \delta\hat{\alpha}'_s]/4,\\
\hat{i}_4 &= q[e^{-i\frac{\pi}{2}}(\alpha_{LO}^* + \delta\hat{\alpha}_{LO}^{'\dag}) - \alpha_s^* - \delta\hat{\alpha}_s^{'\dag}]\times\\&[e^{i\frac{\pi}{2}}(\alpha_{LO} + \delta\hat{\alpha}'_{LO}) - \alpha_s - \delta\hat{\alpha}'_s]/4.
\end{array}
\end{equation}
Here we have absorbed the vacuum noise terms $\delta\hat{\alpha}_v$ into the terms $\delta\hat{\alpha}'$. For simplicity, let us assume that $\alpha_{LO}$ is a real number. To derive the quadratures $\hat{X}$ and $\hat{P}$, the difference of the two photocurrents should be measured
\begin{equation}
\begin{array}{lll} \label{xp}
\hat{\delta i_x} &= i_1 - i_2\\
           &\approx q(\alpha_{LO}\alpha_s^* + \alpha_{LO}\alpha_s + \alpha_{LO}\delta\hat{\alpha}_s^{'\dag} + \alpha_{LO}\delta\hat{\alpha}'_s)/2\\
           &= \frac{q\alpha_{LO}}{2}(\alpha_s+\alpha_s^* + \delta\hat{\alpha}'_s+\delta\hat{\alpha}_s^{'\dag})\\
           &= \frac{q\alpha_{LO}}{2}(X + \delta\hat{X}'),\\
           \\
&\rightarrow \hat{X}_B = \frac{2}{q\alpha_{LO}}\hat{\delta i_x} = X + \delta\hat{X}',\\
\\
\hat{\delta i_p} &= i_3 - i_4\\
           &\approx q(i\alpha_{LO}\alpha_s^* - i\alpha_{LO}\alpha_s + i\alpha_{LO}\delta\hat{\alpha}_s^{'\dag} - i\alpha_{LO}\delta\hat{\alpha}'_s)/2\\
           &= \frac{q\alpha_{LO}}{2}(\frac{\alpha_s-\alpha_s^*}{i} + \frac{\delta\hat{\alpha}'_s-\delta\hat{\alpha}_s^{'\dag}}{i})\\
           &= \frac{q\alpha_{LO}}{2}(P + \delta\hat{P}'),\\
           \\
&\rightarrow \hat{P}_B = \frac{2}{q\alpha_{LO}}\hat{\delta i_p} = P + \delta\hat{P}',\\
\end{array}
\end{equation}
where $X \equiv \alpha_s + \alpha_s^*$ and $P \equiv -i (\alpha_s - \alpha_s^*)$ are the exact quadratures that Bob wants to measure, $\delta\hat{X}' \equiv (\delta\hat{\alpha}_s+\delta\hat{\alpha}_s^{\dag}) + (\delta\hat{\alpha}_v+\delta\hat{\alpha}_v^{\dag}) = \delta\hat{X} + \delta\hat{X}_v$ and $\delta\hat{P}' \equiv -i(\delta\hat{\alpha}_s-\delta\hat{\alpha}_s^{\dag}) -i(\delta\hat{\alpha}_v-\delta\hat{\alpha}_v^{\dag}) = \delta{P} + \delta{P}_v$ are the quantum noises entering into Bob's measurement. Several terms have been neglected above according to the fact that $|\alpha_{LO}| \gg |\alpha_s|$. $\delta\hat{X}$ and $\delta\hat{P}$ satisfy the canonical commutation relation $[\delta\hat{X},\delta\hat{P}] = 2i$, therefore the Heisenberg uncertainty relation $\langle(\delta\hat{X})^2\rangle\langle(\delta\hat{P})^2\rangle = 1$ is derivable~\cite{cv-rev}.

Under the condition that Eve cannot act on the LO (a common assumption in the security proofs~\cite{Sca09}), it is only when the excess noise reaches two times the shot-noise level that Eve can perform an intercept-resend attack on the channel~\cite{cv-sec0}. It is due to the fact that Eve will introduce vacuum noise by using heterodyne detection and consequently, suffer the quantum fluctuations when she reproduces the signal state in a simple intercept-resend attack.

\subsection{Wavelength-dependent fiber beam splitter}\label{WFBS}

In Ref.~\cite{WaveAtt}, we studied the wavelength-dependent property of the fiber beam splitter which is made by the fused biconical taper technology \cite{bs-0}. The fused biconical taper beam splitter is made by closing two or more bare optical fibers, fusing them in a high temperature environment and drawing their two ends at the same time. Subsequently, a specific biconic tapered waveguide structure can be formed in the heating area. The fused biconical taper beam splitter is widely use in the fiber QKD systems because of the feature of low insertion loss, good directivity and low cost.
However, intensity transmission of the fused biconical taper beam splitter is wavelength-dependent, and most types of fused biconical taper beam splitters work only in a limited range of wavelengths (limited bandwidths), where the intensity transmission of the beam splitter can be defined as $T \equiv I_{port1}/(I_{port1}+I_{port2})$, where $I_{port1}$ ($I_{port2}$) is output light intensity from beam splitter's output port 1 (2).
Typical coupling ratio at the center wavelength provides optimal performance, but the intensity transmission varies periodically with wavelength changes. The relationship between wavelength $\lambda$ and the intensity transmission $T$ by using the coupling model is given in Ref.~\cite{bs-1,bs-2}:
\begin{equation}
\begin{array}{lll} \label{bs}
T = F^2 \sin^{2}\Big(\frac{C\lambda^{5/2}w}{F}\Big) \equiv T(\lambda).
\end{array}
\end{equation}
where $F^2$ is the fraction of power coupled, $C\cdot\lambda^{5/2}$ is the coupling coefficient, and $w$ is the heat source width.

\section{Wavelength Attack on a CV-QKD system using heterodyne protocol}\label{Wavelength Attack}

\noindent
\begin{figure}[!h]\center
\resizebox{9cm}{!}{
\includegraphics{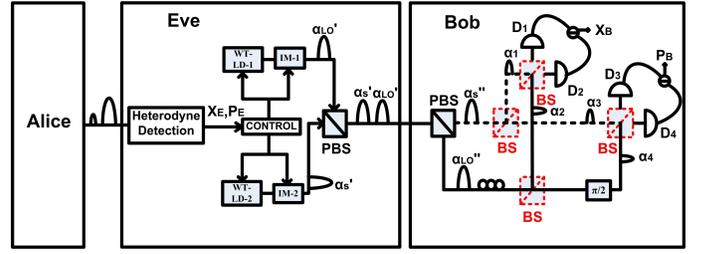}}
\caption{(Color on line)The schematic diagram of the wavelength attack scheme. WT-LD: the wavelength tunable laser diode; IM: the intensity modulator; BS: 50/50 beam splitter. The WT-LD and IM are used in producing fake coherent states with the specific wavelength and amplitude set by the controller. The red (dotted) beam splitters are the ones controlled by Eve. The red beam splitter on the left has transmission $T_1$, while the red beam splitter down the bottom has transmission $T_2$. For simplicity, the 10:90 beam splitter and the generation of $|\alpha_3\rangle$ are not shown.}\label{fig2}
\end{figure}

The basic idea of the wavelength attack is shown in Fig.~\ref{fig2}. Eve intercepts the coherent states sent by Alice. She makes heterodyne measurement of the signal using the LO to achieve the quadrature values $X_E$ and $P_E$. After that, Eve generates and re-sends three coherent states: a fake signal state $|\alpha_s'\rangle$, a fake LO state $|\alpha_{LO}'\rangle$ and together with a ancillary state $|\alpha_3\rangle$. Different from the previous intercept-resend attack, these fake states have different wavelengthes, denoted as $\lambda_1$ (for $|\alpha_s'\rangle$), $\lambda_2$ (for $|\alpha_{LO}'\rangle$) and $\lambda_3$ (for $|\alpha_3\rangle$). According to Eq.~(\ref{bs}), the performance of Bob's beam splitter is dependent on the wavelength of the incoming light. Therefore the fake signal with wavelength $\lambda_1$, the transmission of Bob's beam splitter is determined by the function $T(\lambda_1)$ which is defined in Eq.~(\ref{bs}). Similarly, the intensity transmission of Bob's beam splitter to the fake LO state is determined by $T(\lambda_2)$. In other words, Eve can control Bob's beam splitter by tuning the wavelength of her fake states.

With the help of the wavelength tunable laser diodes and intensity modulators, the wavelength and amplitude of these fake states are carefully chosen to satisfy the following conditions
\begin{equation}
\begin{array}{lll} \label{att}
(i)  & (1-T_3')|\alpha_{3}'|^2 + (1-T_1')|\alpha_{s}'|^2 + (1-T_2')|\alpha_{LO}'|^2\\
     &= 0.1|\alpha_{LO}|^2,\\
(ii)  &(1-T_1)(1-2T_1)T_1'|\alpha_s'|^2 + (1-T_2)(2T_2-1)T_2'|\alpha_{LO}'|^2 \\
       &= \frac{\sqrt{\eta}X_E|\alpha_{LO}|}{2},\\
(iii) &T_1(1-2T_1)T_1'|\alpha_{s}'|^2 + T_2(2T_2-1)T_2'|\alpha_{LO}'|^2 \\
     &= \frac{\sqrt{\eta}P_E|\alpha_{LO}|}{2},
\end{array}
\end{equation}
where $T_i \equiv T(\lambda_i) \in [0,1](i = 1,2), T'_j \equiv T'(\lambda_j) \in [0,1](j=1,2,3)$. Here $\eta$ is the channel transmission efficiency, $|\alpha_s|$ and $|\alpha_{LO}|$ are the amplitudes of the original signal and the LO, respectively, $|\alpha_s'|$, $|\alpha_{LO}'|$ and $\alpha_3'$ are the amplitudes of the fake signal and the fake LO, $T'_j$ are the intensity transmissions of Bob's 10:90 beam splitter (for monitoring the LO light intensity).

Condition (i) makes sure the method of monitoring the LO intensity is invalid to Eve. Here we assume that Bob uses a 10:90 beam splitter to split the total light before being inserted into the PBS~\cite{note}. Because the 10:90 beam splitter is also wavelength-dependent, its intensity transmission can be determined by a function similar to Eq.~(\ref{bs}), which is denoted by $T'(\lambda) \simeq F'^2 \sin^{2}(\frac{C'\lambda^{5/2}w'}{F'})$. Here $|\alpha_3\rangle$ is used for compensating the intensity when $\alpha'_s$ and $\alpha'_{LO}$ are both small. Eve selects an appropriate wavelength $\lambda_3$ such that $T_3 = 0$, therefore the intensity of $|\alpha_3\rangle$ is much lower than the shot noise level and negligible.

As Bob measures the quadratures $\hat{X}_B$ and $\hat{P}_B$ by performing heterodyne detection on the fake signal and the fake LO, conditions (ii) and (iii) make Bob's measurement results coincide with the ones attained by Eve. To see explicitly where these relations come from, see Eqs.~(\ref{eq Appen 1}) and (\ref{eq Appen 2}) in the Appendix. Notice that the fake signal and the fake LO have different wavelengths, and hence, no interference occurs in this detection. The effect of this on the measurement detection is that we no longer have heterodyne detection outputs but rather outputs that are proportional to Eve's measurements. Therefore, the photocurrents recorded by the photo-detectors consist of parts from the signal and the LO. Eve should also make $T'_1|\alpha'_s|^2$ and $T'_2|\alpha'_{LO}|^2$ much smaller than $|\alpha_{LO}|^2$ in order to suppress the shot noise. We are going to prove in Section~\ref{Feasibility Analysis} that the extra noise introduced by Bob's measurement is much lower than the shot-noise level, therefore the total noise can be kept under the alarm threshold. In other words, Eve can safely achieve the key information without being discovered by Alice or Bob.

Finally, we note that as there are limitations on the intensities, conditions (ii) and (iii) may not always be satisfied. However, as the analysis in Appendix.~\ref{appendix 2}, we find that the probability of failing condition (ii) or (iii) is extremely close to zero.

\section{Feasibility Analysis}\label{Feasibility Analysis}

To analyze the feasible of the wavelength attack, we first note that the following assumptions should be satisfied:

(1) This attack is restricted to an all-fiber coherent-state CV-QKD using heterodyne protocol.

(2) All of Bob's beam splitters have the same wavelength dependent property, i.e., their intensity transmissions are all determined by Eq.(\ref{bs}) with the same parameters. This function and the detection efficiencies of Bob's detectors are both known by Eve. Here we assume that the detection efficiencies are wavelength independent for simplicity. In practice, Eve can simply absorb the differences into the light amplitudes modulated by her and the final results are unchanged.

(3) Eve has the ability to replace the quantum channel with a noiseless fiber, and her detectors have high efficiency and negligible excess noise.

Before analyzing the feasibility of the wavelength attack, let us first rapidly review the security analysis of the Gaussian protocols based on coherent states and heterodyne detections under individual attacks. In what follows, we restrict ourselves to Gaussian attacks which are proven optimal~\cite{GarciaPatron2007}.

In the case of Gaussian attacks, the channel connecting Alice and Bob can be fully characterized by its transmission $\eta$, and its excess noise $\epsilon$ above the shot noise level, such that the total noise measured by Bob is $1+\eta\epsilon$ (in shot noise units)~\cite{cv-exp}. Alternatively, one may use the total added noise defined as $\chi \equiv (1-\eta)/\eta + \epsilon$ for convenience.
The secret key rates for Heisenberg-limited individual attack in direct reconciliation and reverse reconciliation are given, respectively, by~\cite{GarciaPatron2007}
\begin{align}\label{R1}
K^{DR} &= \log_2\frac{(1+\chi)[1 + \eta(V+\chi)]}{(1 + \chi V)[1 + \eta(1+\chi)]},
\end{align}
\begin{align}\label{R2}
K^{RR} &= \log_2\frac{V + \eta(1+\chi V)}{\eta(1+\chi V)[1+\eta(1+\chi)]},
\end{align}
where $V = V_A + 1$ is the variance of Alice's modulated state as it was mentioned in Sec.II.A. Note that we use the `Heisenberg-limited attack' rather than the optimal attack~\cite{het-sec,GarciaPatron2007} as such an attack upper bounds Eve's information thereby emphasizing our wavelength attack which can even beat such a stringent attack. From the above formulas, we can see that when V and $\eta$ are settled in practice, the secret key rate is fully determined by $\chi$, which can be precisely estimated from the experimental data~\cite{cv-exp}.

Another important parameter in the security proof is Alice's (Bob's) conditional variance of Bob's (Alice's) measurement $V_{A|B}$ ($V_{B|A}$) in direct reconciliation (reverse reconciliation),which can be thought of as the uncertainty in Alice's (Bob's) estimates of Bob's (Alice's) quadrature measurement result. In the CV-QKD, Alice and Bob use $V_{A|B}$ ($V_{B|A}$) to estimate the shot noise and modulation imperfections~\cite{cv-exp}. $V_{A|B}$ is defined(where both quadratures are symmetrized) as
\begin{equation}
\begin{array}{lll}\label{VAB}
V_{A|B} = \langle X_A^2\rangle - \frac{\langle X_A\hat{X}_B\rangle^2}{\langle\hat{X}_B^2\rangle},
\end{array}
\end{equation}
and similarly, we have $V_{B|A}$ defined as
\begin{equation}
\begin{array}{lll}\label{VBA}
V_{B|A} = \langle \hat{X}_B^2\rangle - \frac{\langle X_A\hat{X}_B\rangle^2}{\langle X_A^2\rangle}.
\end{array}
\end{equation}

We note that $V_{A|B}$($V_{B|A}$) performs a role in CV-QKD protocols similar as the quantum bit error rate in discrete variable QKD protocols, which provide Alice and Bob an intuitive tool to detect the presence of Eve. To clarify this idea, let us first state the relation between $V_{A|B}$($V_{B|A}$) and $\chi$. As the Gaussian character of the channel maintains no matter Eve performs the Gaussian attacks or not, the conditional variance between Alice and Bob, which we will denote as $V_{A|B}^{normal}$ and $V_{B|A}^{normal}$, can be calculated as follows~\cite{GarciaPatron2007}
\begin{align} \label{noise1}
V_{A|B}^{normal}  &= \frac{(V-1)[\eta(\chi+1)+1]}{\eta(V+\chi)+1},\\
V_{B|A}^{normal}  &= \frac{1}{2}[\eta(1+\chi)+1].
\end{align}
Note that there may be a little different from the expressions in~\cite{GarciaPatron2007} due to the differences on the definitions of V.

On the other hand, to make the secret key rate positive, we require that ( according to Eq.~(\ref{R1}) and~(\ref{R2}) )
\begin{align}\label{eq: chi max DR}
\chi < \chi_{max}^{DR} = \frac{\sqrt{4\eta^2+1}-1}{2\eta},
\end{align}
with $\frac{2}{3} < \eta < 1$ for direct reconciliation or
\begin{align}\label{eq: chi max RR}
\chi < \chi_{max}^{RR} = \frac{\sqrt{(\frac{4}{\eta^2}+1)V^2-2V+1}-V-1}{2V},
\end{align}
with $0 < \eta < 1$ for reverse reconciliation, should be satisfied.

Combining Eq.~(\ref{noise1}) with Eq.~(\ref{eq: chi max DR}), we can find that for the sake of deriving a positive secret key rate, the upper bound of $V_{A|B}^{normal}$ yields
\begin{equation}
\begin{array}{lll} \label{max1}
V_{A|B}^{max} &= \frac{(V-1)(\sqrt{4\eta^2+1}+2\eta+1)}{\sqrt{4\eta^2+1}+2\eta V + 1}.
\end{array}
\end{equation}
In other words, if $V_{A|B}$ is smaller than this threshold, the heterodyne protocol in direct reconciliation is considered to be secure. Similarly, the upper bound of $V_{B|A}^{normal}$ is derived to be
\begin{equation}
\begin{array}{lll} \label{max2}
V_{B|A}^{max} &= \frac{\sqrt{(4+\eta^2)V^2-2\eta^2V+\eta^2}+(\eta+2)V-\eta}{4V}.
\end{array}
\end{equation}
And the heterodyne protocol in reverse reconciliation is considered to be secure if $V_{B|A}$ is smaller than this threshold.

For these reasons, we can prove our attack feasible by showing that Eve can make $V_{A|B} < V_{A|B}^{max}$ (in the direct reconciliation protocol) and $V_{B|A} < V_{B|A}^{max}$ (in the reverse reconciliation protocol) when she is performing the wavelength attack.


%
%

\subsection{Eve's Wavelength Attack}\label{Eve's Wavelength Attack}

When Eve performs the wavelength attack, with channel noise, from a real value $X_A$ chosen by Alice to the measurement result $\hat{X}_B$ achieved by Bob is listed as follows (we write down the quadrature $\hat{X}$ only since the other quadrature $\hat{P}$ can be presented in the similar way)
\begin{align}\nonumber
X_A  \rightarrow \hat{X}_A &= X_A + \hat{N}_A \\\label{eq: presence of Eve}
    \rightarrow \hat{X}_E &= \frac{1}{\sqrt2}(\hat{X}_A + \hat{N}_E)\\\nonumber
    \rightarrow \hat{X}_B &= \sqrt{\eta} \hat{X}_E + \hat{N}_{B},
\end{align}
%
where $\hat{N}_E$ represents the vacuum noise in Eve's heterodyne detection whose variance is normalized to $1$, and $\hat{N}_B$ is the vacuum noise introduced by the heterodyne detection. The variance of each of the terms is given by: $V_E = \frac{1}{2}(V + 1)$ and $V_B = \eta V_E + V_{NB}$. Here $V_{NB}$ can then be considered as Eve's conditional variance of Bob's measurement result. In Appendix~\ref{appendix 1}, we derive the value of $V_{NB}$ and show that it is smaller than 0.13. We are now ready to derive the conditional variances under Eve's attack, which are denoted as $V_{A|B}^{attack}$ and $V_{B|A}^{attack}$.

\subsubsection{$V_{A|B}^{attack}$ in direct reconciliation}\label{VaDR}

\begin{figure}[!h]\center
\resizebox{9cm}{!}{
\includegraphics{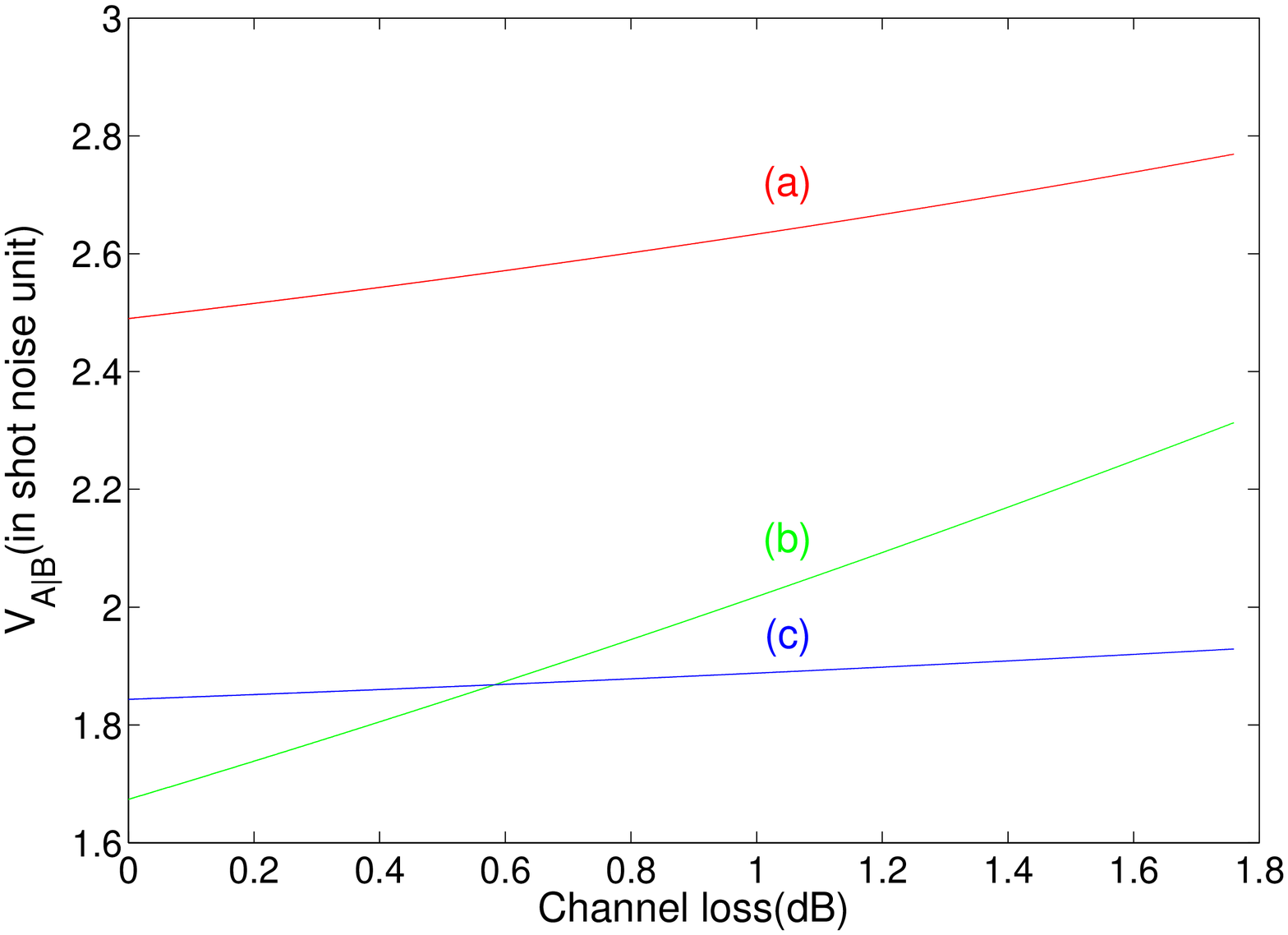}}
\caption{(Color on line)In direct reconciliation, the relation between the channel loss and the conditional variance $V_{A|B}$ in three cases: (a) the maximum tolerable value $V_{A|B}^{max}$. (b) the value of $V_{A|B}^{normal}$ and (c) the value of $V_{A|B}^{attack}$. See main text for details. The curves are plotted for experimentally realistic values, $V = 11$ and $\epsilon = 0.01$. We can see that $V_{A|B}^{attack}$ is always lower than $V_{A|B}^{max}$ and lower than $V_{A|B}^{normal}$ when the channel loss is larger than $0.58$~dB at which point the key between Alice and Bob is no longer secure.}\label{fig3}
\end{figure}

According to the definition of $V_{A|B}$ in Eq.~(\ref{VAB}), the value of $V_{A|B}^{attack}$ can be computed as follows
\begin{equation}
\begin{array}{lll} \label{VABa}
V_{A|B}^{attack}
          &= \frac{2(V_{NB}+\eta)(V-1)}{2V_{NB}+\eta(V+1)}.
\end{array}
\end{equation}
Combining with Eq.~(\ref{VNBx},\ref{VNBp}) and the discussions above, we can estimate that the value of $V_{A|B}^{attack}$ is not larger than 1.9. As shown in Fig.~\ref{fig3} (where we set $V = 11$ and $\epsilon=0.01$~\cite{het-sec}), $V_{A|B}^{attack}$ is always lower than $V_{A|B}^{max}$, so that Alice and Bob can never discover the eavesdropper under this attack.
Besides, one should notice that $V_{A|B}^{attack}$ is always lower than the normal level when the channel loss is larger than $0.58$~dB, therefore Eve should increase the deviations on purpose to make $V_{A|B}^{attack}$ close to $V_{A|B}^{normal}$ in order to avoid suspicion.

\subsubsection{$V_{B|A}^{attack}$ in reverse reconciliation}\label{VaRR}

\begin{figure}[!h]\center
\resizebox{9cm}{!}{
\includegraphics{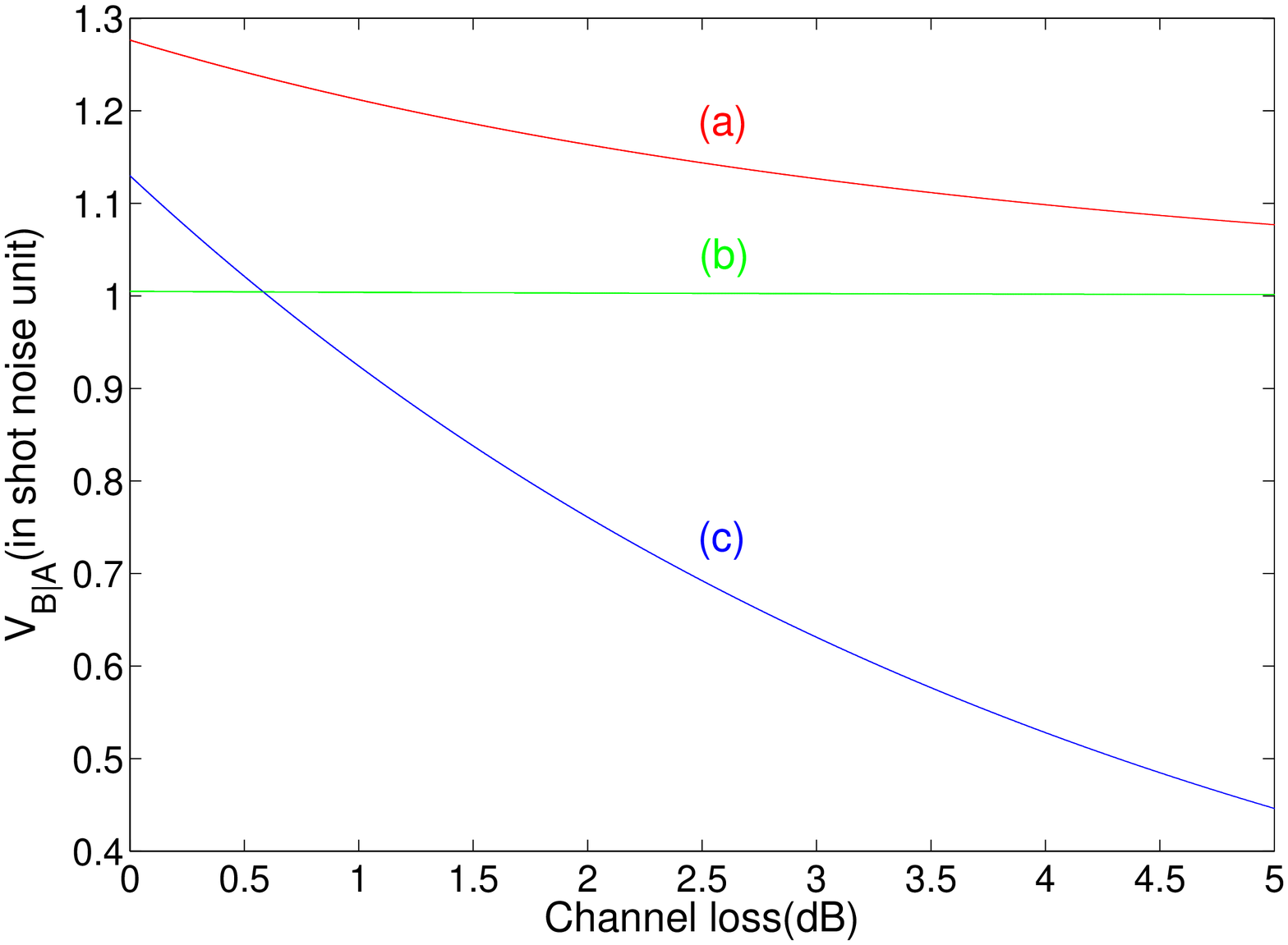}}
\caption{(Color on line)In reverse reconciliation, the relation between the channel loss and the conditional variance $V_{B|A}$ in three cases: (a) the maximum tolerable value $V_{B|A}^{max}$. (b) the value of $V_{B|A}^{normal}$ and (c) the value of $V_{B|A}^{attack}$. See main text for details. The curves are plotted for experimentally realistic values, $V=11$ and $\epsilon=0.01$. We can see that $V_{B|A}^{attack}$ is always lower than $V_{B|A}^{max}$ and lower than $V_{B|A}^{normal}$ when the channel loss is greater than $0.58$~dB, again leading to an insecure key.}\label{fig4}
\end{figure}

In reverse reconciliation, using Eq.~(\ref{VBA}) with Eq.~(\ref{eq: presence of Eve}), the value of $V_{B|A}^{attack}$ can be computed as
\begin{equation}
\begin{array}{lll} \label{VBAa}
V_{B|A}^{attack} = \eta + V_{NB}.
\end{array}
\end{equation}
Combining with Eq.~(\ref{VNBx},\ref{VNBp}) and the discussions above, we can estimate that the value of $V_{B|A}^{attack}$ is never larger than $\eta+0.13$. As shown in Fig.~\ref{fig4} (where again we have set $V = 11$ and $\epsilon=0.01$), it is always lower than the value of $V_{B|A}^{max}$ so that Alice and Bob can never discover the eavesdropper under such an attack.
Besides, one should notice that $V_{B|A}^{attack}$ is lower than the $V_{B|A}^{normal}$ when the channel loss is larger than $0.58$~dB. Hence, Eve should increase the deviations to make $V_{A|B}^{attack}$ close to $V_{A|B}^{normal}$ in order to avoid suspicions.

\section{Discussions and Conclusion}\label{conclusion}

There are two points about the wavelength attack that should be remarked:

\begin{enumerate}
  \item As shown in Fig.~\ref{fig3} and Fig.~\ref{fig4}, $V_{A|B}^{attack}$ and $V_{B|A}^{attack}$ are lower than $V_{A|B}^{normal}$ and $V_{B|A}^{normal}$ respectively when $\eta < 0.88$. It is impossible when the protocol works normally, therefore Eve should add extra noise on her measurement result to increase $V_{A|B}^{attack}$ and $V_{B|A}^{attack}$. So that perfect heterodyne detection is not necessary for Eve. In other words, assumption (3) listed in Section~\ref{Feasibility Analysis} can be compromised.
  \item In theory, the wavelength attack cannot be avoided by adding wavelength filter before the monitoring detector, because Eve can simply increase the input light intensity~\cite{WaveAtt}. To make this method work, Bob should randomly choose to add or not to add a wavelength filter before the monitoring detector and observe the differences.
\end{enumerate}
%


Finally, we note that a commercial CV-QKD system, as sold by~\cite{securenet}, currently uses a wavelength-dependent beam splitter. Although, it does not fall into the regime studied in this paper because it uses homodyne detection rather than heterodyne detection. However, our results show that if one were going to use heterodyne detection with a commercial QKD unit, then the precautions mentioned here would need to be taken. Furthermore, possible quantum hacking opportunities with homodyne detection and wavelength-dependent beam splitters warrant further investigation.

In conclusion, we have proposed a new type of realistic quantum hacking attack, namely the wavelength attack, on continuous-variable QKD systems using heterodyne detection. If Alice and Bob don't take the necessary precautions for such an attack, the final secret key is in principle, totally insecure as Eve can obtain all the information about the final key. This is different from the equal-amplitude attack proposed in Ref.~\cite{testlo} as in the wavelength attack, Eve has the ability to control Bob's beam splitter and therefore the suggestion of testing the total intensity in Ref.~\cite{testlo} would not prevent such an attack from occurring. To close such a loophole in practical CV-QKD systems, it is simply enough for Bob to randomly add a wavelength filter before his detection.

\emph{Added note: To suppress the shot noise, Eve can also apply squeezed state instead of coherent state to generate the fake pulses. In this case, the constraint about maximum fake pulse intensity can be loosed. We would like to thank Dr. Bing Qi for providing this idea to us.}

\subsection{Acknowledgement}

This work was supported by the National Basic Research Program of China (Grants No. 2011CBA00200 and No. 2011CB921200), National Natural Science Foundation of China (Grants No. 60921091 and No. 61101137). C.~W. acknowledges support from the Ontario postdoctoral fellowship program, CQIQC postdoctoral fellowship program, CIFAR, Canada Research Chair program,
NSERC, and QuantumWorks.


\appendix

\section{Achievable $X_E$ and $P_E$}\label{appendix 2}
We estimate the achievable range of $X_E$ and $P_E$ in this appendix. Before the analysis, let us first rewrite Eq.~(\ref{bs}) as follow
\begin{align} \label{bs2}
T(\lambda) &= F^2 \sin^{2}\Big(\frac {Cw}{F}\lambda^{5/2}\Big) = \sin^2(AX),
\end{align}
where $A=\frac{Cw}{F}$ and $X = \lambda^{5/2}$, here we set F=1 for simplicity. For the $50:50$ BS, $T(\lambda_0) = \sin^2(AX_0) = 0.5$ where $\lambda_0 =$ 1550nm, hence $AX_0 = \arcsin(\sqrt{0.5})$. For other wavelengths, $AX = \arcsin(\sqrt{T(\lambda)})$ and we can get $X = \frac{\arcsin(\sqrt{T(\lambda)})}{\arcsin(0.5)}X_0$.

For the $10:90$ BS, we similarly rewrite its transmission as $T'(\lambda) = \sin^2(BX)$ and easily derive that $BX_0 = \arcsin(\sqrt{0.9})$. Therefore,
\begin{align} \label{app2}
T'(\lambda) = sin^2(\frac{arcsin(\sqrt{T(\lambda)})}{arcsin\sqrt{0.5}}arcsin\sqrt{0.9}).
\end{align}

Moreover, as it is mentioned in Sec.~\ref{Eve's Wavelength Attack}, for suppressing the shot noise we should make $|\alpha''_s| \equiv T'_1|\alpha'_s|^2$ and $|\alpha''_{LO}| \equiv T'_2|\alpha'_{LO}|^2$ much smaller than $|\alpha_{LO}|^2$. In a practical CV QKD system, the LO pulse arrived at Bob's side typically includes more than $10^8$ photons\cite{Jouguet2012}. For this reason, we constrain the maximum value of both $|\alpha''_s|^2$ and $|\alpha''_{LO}|^2$ to be $10^6 \leq 10^{-2}|\alpha_{LO}|^2$. On the other hand, to guarantee condition (i), Eve should also make $(1-T'_1)|\alpha'_s|^2 $ and $(1-T'_2)|\alpha'_{LO}|^2 $ not larger than $5\times10^6$. We then get the following maximum value constrains on the fake state intensities:
\begin{equation}
\begin{array}{lll} \label{intmax}
|\alpha''_s|^2 \leq Min\{10^6, \frac{5T'_1}{1-T'_1}10^6\},\\
|\alpha''_{LO}|^2 \leq Min\{10^6, \frac{5T'_2}{1-T'_2}10^6\}.
\end{array}
\end{equation}
From condition (ii) and (iii), we can get
\begin{equation}
\begin{array}{lll}\label{xepe}
\sqrt{\eta}X_E &= \frac{2[(1-T_1)(1-2T_1)|\alpha''_s|^2 + (1-T_2)(2T_2-1)|\alpha''_{LO}|^2]}{|\alpha_{LO}|},\\
\sqrt{\eta}P_E &= \frac{2[T_1(1-2T_1)|\alpha''_s|^2 + T_2(2T_2-1)|\alpha''_{LO}|^2]}{|\alpha_{LO}|}.
\end{array}
\end{equation}
Combining Eq.~(\ref{app2}), (\ref{intmax}) and (\ref{xp}), now we have enough information to derive the achievable value range of $X_E$ and $P_E$ by analytical calculations or numerical simulations. Either of these methods shows that $(\sqrt{\eta}X_E, \sqrt{\eta}P_E)$ satisfying $\eta|X_E|^2 + \eta|P_E|^2 < 20$ are always achievable. To see how high the probability of $|X_E|$(or $|P_E|$) $> 20$ is, we can apply the error integral function erfc$(x)=\frac{2}{\sqrt{\pi}}\int^{\infty}_{x}e^{-x^2}dx$ and get $P[|\sqrt{\eta}X_E|>20$ or $|\sqrt{\eta}P_E|>20] =$ erfc$(\frac{20}{\sqrt{2V}})$, where V is the variance of $X_E$ and $P_E$ chosen by Gaussian distribution\cite{book2}. For an experimentally realistic value V=11, we get erfc$(\frac{20}{\sqrt{22}}) = 1.637\times 10^{-9} \approx 0$, which concludes our claim in Sec.~\ref{Wavelength Attack}. When $X_E$ or $P_E$ is out of reach, Eve can simply turn to perform the original intercept-resend strategy. The extra noise it involves is 1 (shot noise unit) times this extremely low probability, which is negligible.

\section{Derivation of $V_{NB}$}\label{appendix 1}

To derive $V_{NB}$, let us start from the generation of Eve's fake states. As we have described in Section~\ref{Wavelength Attack}, Eve generates the fake signal state and the fake LO state according to her measurement results and sends them to Bob. These fake states can be described by the following operators
\begin{equation}
\begin{array}{lll}
\hat{\alpha}'_s &= \alpha'_s + \delta\hat{a}'_{s},\\
\hat{\alpha}'_{LO} &= \alpha'_{LO} + \delta\hat{a}'_{LO}.
\end{array}
\end{equation}
Where complex numbers $\alpha'_s$ and $\alpha'_{LO}$ are the amplitudes and $\delta\hat{a}'_{s}$ and $\delta\hat{a}'_{LO}$ represent the fluctuations of the amplitudes as discussed in Section~\ref{HDP}. Similarly, $\langle(\delta\hat{X}'_k)^2\rangle = \langle(\delta\hat{P}'_k)^2\rangle = 1$, where $\delta\hat{X}_k = \delta\hat{a}'_k + \delta\hat{a}_k^{'\dag}$ and $\delta\hat{P}_k  =-i(\delta\hat{a}'_k - \delta\hat{a}_k^{'\dag})$, k = s, LO. After the (original) $10:90$ beam splitter, they are turned to be
\begin{equation}
\begin{array}{lll}
\hat{\alpha}''_s &= \sqrt{T'_1}\hat{\alpha}'_s + \sqrt{1-T'_1}\delta\hat{a}_{v1}\\
                 &= \alpha''_s + \delta\hat{a}''_{s},\\
\hat{\alpha}''_{LO} &= \sqrt{T'_2}\hat{\alpha}'_{LO} + \sqrt{1-T'_2}\delta\hat{a}'_{v2}\\
                    &= \alpha''_{LO} + \delta\hat{a}''_{LO},
\end{array}
\end{equation}
where $\delta\hat{a}_{v1}$, $\delta\hat{a}_{v2}$ are the vacuum noises that interfere with the fake signal and the fake LO, respectively, at the beam splitter. $\alpha''_s \equiv \sqrt{T'_1}\alpha'_s$, $\delta\hat{a}''_{s} \equiv \sqrt{T'_1}\delta\hat{a}'_{s} + \sqrt{1-T'_1}\delta\hat{a}'_{v1}$ and similar to LO.

Bob performs heterodyne detection on these fake states. According to Eq.~(\ref{bs}), Bob's beam splitter has different intensity transmissions for $\hat{\alpha}''_s$ and $\hat{\alpha}''_{LO}$ because of their different wavelengths, denoted as $T_1$ and $T_2$. After passing the first set of beam splitters, $\hat{\alpha}''_s$ is separated into $\hat{\alpha}_1$ and $\hat{\alpha}_3$, while $\hat{\alpha}''_{LO}$ is separated into $\hat{\alpha}_2$ and $\hat{\alpha}_4$ (cf. Fig.~\ref{fig2}), which can be expressed as follows
\begin{equation}
\begin{array}{lll}
\hat{\alpha}_1 = \sqrt{1-T_1}\hat{\alpha}_s'' + \sqrt{T_1}\delta\hat{a}'_{v1},\\
\hat{\alpha}_2 = \sqrt{1-T_2}\hat{\alpha}_{LO}'' + \sqrt{T_2}\delta\hat{a}'_{v2},\\
\hat{\alpha}_3 = \sqrt{T_1}\hat{\alpha}_s'' - \sqrt{1-T_1}\delta\hat{a}'_{v1},\\
\hat{\alpha}_4 = e^{i\frac{\pi}{2}}(\sqrt{T_2}\hat{\alpha}_{LO}'' - \sqrt{1-T_2}\delta\hat{a}'_{v2}),
\end{array}
\end{equation}
%
To simplify the symbols, let us define
$\delta\hat{\alpha}_1 \equiv \sqrt{1-T_1}\delta\hat{a}''_s + \sqrt{T_1}\delta\hat{\alpha}'_{v1}$,
$\delta\hat{\alpha}_2 \equiv \sqrt{1-T_2}\delta\hat{a}''_{LO} + \sqrt{T_2}\delta\hat{\alpha}'_{v2}$,
$\delta\hat{\alpha}_3 \equiv \sqrt{T_1}\delta\hat{a}''_s - \sqrt{1-T_1}\delta\hat{\alpha}'_{v1}$ and
$\delta\hat{\alpha}_4 \equiv \sqrt{T_2}\delta\hat{a}''_{LO} - \sqrt{1-T_2}\delta\hat{\alpha}'_{v2}$.
Furthermore, we define the quadratures of $\delta\hat{\alpha}_k$ by
$\delta\hat{X}_{k} = \delta\hat{\alpha}_k + \delta\hat{\alpha}_k^{\dag}$ and $\delta\hat{P}_{k} = -i(\delta\hat{\alpha}_k - \delta\hat{\alpha}_k^{\dag})$ where $k = 1, 2, 3, 4$. Finally, after combining at the second set of beam splitters, the electromagnetic fields arrive at the four detectors can be written as
\begin{equation}
\begin{array}{lll}
\hat{b}_1 = \sqrt{1-T_1}\hat{\alpha}_1 + \sqrt{T_1}\delta\hat{\alpha}''_{v1} + \sqrt{T_2}\hat{\alpha}_2 + \sqrt{1-T_2}\delta\hat{\alpha}''_{v2},\\
\hat{b}_2 = \sqrt{T_1}\hat{\alpha}_1 - \sqrt{1-T_1}\delta\hat{\alpha}''_{v1} - \sqrt{1-T_2}\hat{\alpha}_2 + \sqrt{T_2}\delta\hat{\alpha}''_{v2},\\
\hat{b}_3 = \sqrt{1-T_1}\hat{\alpha}_3 + \sqrt{T_1}\delta\hat{\alpha}''_{v3} + \sqrt{T_2}\hat{\alpha}_4 + \sqrt{1-T_2}\delta\hat{\alpha}''_{v4},\\
\hat{b}_4 = \sqrt{T_1}\hat{\alpha}_3 - \sqrt{1-T_1}\delta\hat{\alpha}''_{v3} - \sqrt{1-T_2}\hat{\alpha}_4 + \sqrt{T_2}\delta\hat{\alpha}''_{v4}.
\end{array}
\end{equation}
where the photocurrents are given by $\hat{i}_k = q\hat{b}_k^{\dag}\hat{b}_k$. Bob's quadrature measurement results are then derived from the difference in photocurrents, using the method in Section~\ref{HDP}. Firstly, for detectors $D1$ and $D2$, we have
\begin{widetext}
\begin{equation}\label{eq Appen 1}
\begin{array}{lll}
\hat{i}_x &=\hat{i}_1 -\hat{i}_2\\
          &=q(\hat{b}_1^{\dag}\hat{b}_1 - \hat{b}_2^{\dag}\hat{b}_2)\\
          &=q\{(1-2T_1)[(1-T_1)|\alpha_s''|^2 + \sqrt{1-T_1}(\alpha_s''^{*}\delta\hat{\alpha}_1 + \alpha_s''\delta\hat{\alpha}_1^{\dag})]\\
          &+ (2T_2-1)[(1-T_2)|\alpha_{LO}''|^2 + \sqrt{1-T_2}(\alpha_{LO}''^{*}\delta\hat{\alpha}_2 + \alpha_{LO}''\delta\hat{\alpha}_2^{\dag})]\\
          &+ 2\sqrt{T_1}(1-T_1)(\alpha_s^{''*}\delta\hat{\alpha}''_{v1} + \alpha''_s\delta\hat{\alpha}^{''\dag}_{v1}) + 2\sqrt{T_2}(1-T_2)(\alpha_{LO}^{''*}\delta\hat{\alpha}''_{v2} + \alpha''_{LO}\delta\hat{\alpha}^{''\dag}_{v2})\\
          &+\sqrt{(1-T_1)T_2}[\sqrt{(1-T_1)(1-T_2)}(\alpha_s^{''*}\alpha_{LO}'' + \alpha_s''\alpha_{LO}^{''*})+\sqrt{1-T_1}(\alpha_s''^*\delta\hat{\alpha}_2 + \alpha_s''\delta\hat{\alpha}_2^{\dag}) + \sqrt{1-T_2}(\alpha_{LO}''\delta\hat{\alpha}_1^{\dag} + \alpha_{LO}''^*\hat{\alpha}_1)]\\
          &- \sqrt{(1-T_2)T_1}[\sqrt{(1-T_1)(1-T_2)}(\alpha_s''^*\alpha_{LO}'' + \alpha_s''\alpha_{LO}^{''*}) + \sqrt{1-T_1}(\alpha_s''^*\delta\hat{\alpha}_2 + \alpha_s''\delta\hat{\alpha}_2^{\dag}) + \sqrt{1-T_2}(\alpha_{LO}''\delta\hat{\alpha}_1^{\dag} + \alpha_{LO}''^*\hat{\alpha}_1)]\\
          &+ (\sqrt{T_1T_2(1-T_2)}-\sqrt{(1-T_1)(1-T_2)^2})(\alpha''_{LO}\delta\hat{\alpha}^{''\dag}_{v1}+\alpha^{''*}_{LO}\delta\hat{\alpha}''_{v1})\\
          &+(\sqrt{(1-T_1)^2(1-T_2)}-\sqrt{T_1T_2(1-T_1)})(\alpha_s''\delta\hat{\alpha}^{''\dag}_{v2} +\alpha^{''*}_s\delta\hat{\alpha}''_{v2}) \}.
\end{array}
\end{equation}
Where the terms $\delta\hat{\alpha}^{\dag}\delta\hat{\alpha}$ are already neglected. Note that $\hat{\alpha}'_{LO}$ and $\hat{\alpha}'_s$ have different frequencies, therefore any terms not containing the product of the same frequencies vanish during the measurement. The remaining terms compose the measurement result of $\hat{i}_x$
\begin{equation}\label{eq Appen 1}
\begin{array}{lll}
\hat{i}_x &= q[(1-T_1)(1-2T_1)|\alpha_s''|^2 + (1-2T_1)\sqrt{1-T_1}(\alpha_s''^{*}\delta\hat{\alpha}_1 + \alpha_s''\delta\hat{\alpha}_1^{\dag})\\
          &+ (1-T_2)(2T_2-1)|\alpha_{LO}''|^2 + (2T_2-1)\sqrt{1-T_2}(\alpha_{LO}''^{*}\delta\hat{\alpha}_2 + \alpha_{LO}''\delta\hat{\alpha}_2^{\dag})\\
          &+ 2\sqrt{T_1}(1-T_1)(\alpha_s''^{*}\delta\hat{\alpha}''_{v1} + \alpha_s''\delta\hat{\alpha}''^{\dag}_{v1}) + 2\sqrt{T_2}(1-T_2)(\alpha_{LO}''^{*}\delta\hat{\alpha}'_{v2} + \alpha_{LO}''\delta\hat{\alpha}''^{\dag}_{v2})].
\end{array}
\end{equation}
Similarly, we get the measurement result of $\hat{i}_p$ as:
\begin{equation}
\begin{array}{lll}\label{eq Appen 2}
\hat{i}_p &= \hat{i}_3 -\hat{i}_4 = q(\hat{b}_3^{\dag}\hat{b}_3 - \hat{b}_4^{\dag}\hat{b}_4)\\
          &= q[T_1(1-2T_1)|\alpha_s''|^2 + (1-2T_1)\sqrt{T_1}(\alpha_s''^{*}\delta\hat{\alpha}_3 + \alpha_s''\delta\hat{\alpha}_3^{\dag})\\
          &+ T_2(2T_2-1)|\alpha_{LO}''|^2 + (2T_2-1)\sqrt{T_2}(\alpha_{LO}''^{*}\delta\hat{\alpha}_4 + \alpha_{LO}''\delta\hat{\alpha}_4^{\dag})\\
          &+ 2T_1\sqrt{1-T_1}(\alpha''^*_s\delta\alpha''_{v3}+\alpha''_s\delta\alpha''^{\dag}_{v3}) + i2T_2\sqrt{1-T_2}(\alpha''_{LO}\delta\alpha''^{\dag}_{v4}-\alpha''^*_{LO}\delta\alpha''_{v4})].
\end{array}
\end{equation}
Notice that the squared modulus terms in the last two equations are what helped us derive the set of conditions in Eq.~(\ref{att}). The measurement results corresponding to Bob's quadratures $\hat{X}_B$ and $\hat{P}_B$ are then calculated using Eq.~(\ref{xp}):
\begin{equation}
\begin{array}{lll}
\hat{X}_B &= \frac{2\hat{i}_x}{q|\alpha_{LO}|} \\
    &= \frac{2[(1-T_1)(1-2T_1)|\alpha_s''|^2 + (1-T_2)(2T_2-1)|\alpha_{LO}''|^2]}{|\alpha_{LO}|}\\
       &+ \frac{2[(1-2T_1)\sqrt{1-T_1}(\alpha_s''^{*}\delta\hat{\alpha}_1 + \alpha_s''\delta\hat{\alpha}_1^{\dag}) +  (2T_2-1)\sqrt{1-T_2}(\alpha_{LO}''^{*}\delta\hat{\alpha}_2 + \alpha_{LO}''\delta\hat{\alpha}_2^{\dag})]}{|\alpha_{LO}|}\\
       &+ \frac{4[\sqrt{T_1}(1-T_1)(\alpha_s''^{*}\delta\hat{\alpha}''_{v1} + \alpha_s''\delta\hat{\alpha}_{v1}^{''\dag}) + \sqrt{T_2}(1-T_2)(\alpha_{LO}''^{*}\delta\hat{\alpha}''_{v2} + \alpha_{LO}''\delta\hat{\alpha}_{v2}^{''\dag})]}{|\alpha_{LO}|}\\
    &= \sqrt{\eta}X_E + \hat{X}_{NB}, \\
\hat{P}_B &= \frac{2\hat{i}_x}{q|\alpha_{LO}|} \\
    &= \frac{2[T_1(1-2T_1)|\alpha_s''|^2 +  T_2(2T_2-1)|\alpha_{LO}''|^2]}{|\alpha_{LO}|}\\
    &+ \frac{2[(1-2T_1)\sqrt{T_1}(\alpha_s''^{*}\delta\hat{\alpha}_3 + \alpha_s''\delta\hat{\alpha}_3^{\dag})
    + (2T_2-1)\sqrt{T_2}(\alpha_{LO}''^{*}\delta\hat{\alpha}_4 + \alpha_{LO}''\delta\hat{\alpha}_4^{\dag})]}{|\alpha_{LO}|}\\
    &+ \frac{4[T_1\sqrt{1-T_1}(\alpha''^*_s\delta\alpha''_{v3}+\alpha''_s\delta\alpha''^{\dag}_{v3}) + iT_2\sqrt{1-T_2}(\alpha''_{LO}\delta\alpha''^{\dag}_{v4}-\alpha''^*_{LO}\delta\alpha''_{v4})]}{|\alpha_{LO}|}\\
    &= \sqrt{\eta}P_E + \hat{P}_{NB}.
\end{array}
\end{equation}
where we have used conditions (ii) and (iii) from Eq.~(\ref{att}).
Let $\alpha''_{LO}$ and $\alpha''_s$ be real,
we then get the following inequalities:
\begin{equation}
\begin{array}{lll}\label{V}
\hat{X}_{NB} &= \frac{2[(1-2T_1)\sqrt{1-T_1}\alpha_s''(\delta\hat{\alpha}_1 + \delta\hat{\alpha}_1^{\dag}) +  (2T_2-1)\sqrt{1-T_2}\alpha_{LO}''(\delta\hat{\alpha}_2 + \delta\hat{\alpha}_2^{\dag})]}{|\alpha_{LO}|}\\
             &+ \frac{4[\sqrt{T_1}(1-T_1)\alpha_s''(\delta\hat{\alpha}''_{v1} + \delta\hat{\alpha}_{v1}''^{\dag}) + \sqrt{T_2}(1-T_2)\alpha_{LO}''(\delta\hat{\alpha}''_{v2} + \delta\hat{\alpha}_{v2}^{''\dag})]}{|\alpha_{LO}|}\\
\hat{P}_{NB} &= \frac{2[\sqrt{T_1}(1-2T_1)\alpha_s''(\delta\hat{\alpha}_3 + \delta\hat{\alpha}_3^{\dag}) +  \sqrt{T_2}(2T_2-1)\alpha_{LO}''(\delta\hat{\alpha}_2 + \delta\hat{\alpha}_4^{\dag})]}{|\alpha_{LO}|}\\
             &+ \frac{4[T_1\sqrt{1-T_1}\alpha''_s(\delta\alpha''_{v3} + \delta\alpha''^{\dag}_{v3}) + iT_2\sqrt{1-T_2}\alpha''_{LO}(\delta\alpha''^{\dag}_{v4} - \delta\alpha''_{v4})]}{|\alpha_{LO}|}\\
\end{array}
\end{equation}
\end{widetext}
%
%
Therefore,
\begin{equation}
\begin{array}{lll}\label{VNBx}
V_{NB,x} &= \langle(\hat{X}_{NB})^2\rangle \\
         &= \frac{4[\langle(1-2T_1)^2(1-T_1)\alpha_s^{''2}\delta X_1^2\rangle + \langle(2T_2-1)^2(1-T_2)\alpha_{LO}^{''2}\delta X_2^2\rangle]}{|\alpha_{LO}|^2}\\
         &+ \frac{16[\langle T_1(1-T_1)^2\alpha_s^{''2}\delta X_{v1}^{''2}\rangle + \langle T_2(1-T_2)^2\alpha_{LO}^{''2}\delta X_{v2}^{''2}\rangle]}{|\alpha_{LO}|^2}\\
         &< 13\times\frac{max\{|\alpha''_{s}|^2,|\alpha''_{LO}|^2\}}{|\alpha_{LO}|^2} = 0.13,
\end{array}
\end{equation}
\begin{equation}
\begin{array}{lll}\label{VNBp}
V_{NB,p} &= \langle(\hat{P}_{NB})^2\rangle \\
         &= \frac{4[\langle(1-2T_1)^2T_1\alpha_s^{''2}\delta X_3^2\rangle + \langle(2T_2-1)^2T_2\alpha_{LO}^{''2}\delta X_4^2\rangle]}{|\alpha_{LO}|^2}\\
         &+ \frac{16[\langle T_1^2(1-T_1)\alpha_s^{''2}\delta X_{v3}^{''2}\rangle + \langle T_2^2(1-T_2)\alpha_{LO}^{''2}\delta X_{v4}^{''2}\rangle]}{|\alpha_{LO}|^2}\\
         &< 13\times\frac{max\{|\alpha''_{s}|^2,|\alpha''_{LO}|^2\}}{|\alpha_{LO}|^2} = 0.13.
\end{array}
\end{equation}
Here we use the facts that the maximum values of $(1-2T)^2(1-T)$, $(1-2T)^2T$, $T(1-T)^2$ and $T^2(1-T)$ are 1, 1, $\frac{4}{27}$ and $\frac{4}{27}$ respectively, $\langle\delta X^2\rangle = \langle\delta P^2\rangle = 1$, and the constrain of $max\{|\alpha''_{s}|^2,|\alpha''_{LO}|^2\} < 10^{-2} |\alpha_{LO}|^2$(see Appendix~\ref{appendix 2}).


\begin{thebibliography}{10}
\bibitem{Sca09} V.~Scarani, H.~Bechmann-Pasquinucci, N.~J.~Cerf, M.~Du\v{s}ek, N.~Lutkenhaus, and M.~Peev,
Rev.~Mod.~Phys. \textbf{81}, 1301 (2009).

\bibitem{qkd} N.~Gisin, G.~Ribordy, W.~Tittel, and H.~Zbinden, Rev.~Mod.~Phys. \textbf{74}, 145 (2002).

\bibitem{Qi2007} B.~Qi, C.~-H.~F.~Fung, H.~-K.~Lo, and X.~Ma, Quant.~Inf.~Comp. \textbf{7}, 7382 (2007).

\bibitem{Zhao2008} Y.~Zhao, C.~-H.~F.~Fung, B.~Qi, C.~Chen, and H.~-K.~Lo, Phys.~Rev.~A \textbf{78}, 042333 (2008).

\bibitem{Lydersen2010} L.~Lydersen, C.~Wiechers, C.~Wittmann, D.~Elser, J.~Skaar, and V.~Makarov, Optics Express. \textbf{8}, 27938-27954 (2010).

\bibitem{Lars} L.~Lydersen, C.~Wiechers, C.~Wittmann, D.~Elser, J.~Skaar, and V.~Makarov,
Nature Photonics. \textbf{4}, 686-689 (2010).


\bibitem{Xu} F.~-H. Xu, B.~Qi, and H.~-K.~Lo, New J.Phys. \textbf{12}, 113026 (2010).

\bibitem{Gerhardt2011} I.~Gerhardt,	 Q.~Liu, A.~Lamas-Linares,	 J.~Skaar, C.~Kurtsiefer, and V.~Makarov, Nature Comm. \textbf{2}, 349 (2011).

\bibitem{Wiechers2011} C~.Wiechers, L.~Lydersen, C.~Wittmann, D.~Elser, J.~Skaar, C.~Marquardt, V.~Makarov, and G.~Leuchs, New~J.~Phys. \textbf{13}, 013043 (2011).

\bibitem{Weier} H.~Weier, H.~Krauss, M.~Rau, M.~F\"{u}erst, S.~Nauerth, and H.~Weinfurter,
New~J.~Phys. \textbf{13}, 073024 (2011).

\bibitem{Jain} N.~Jain, C.~Wittmann, L.~Lydersen, C.~Wiechers, D.~Elser, C.~Marquardt, V.~Makarov, and G.~Leuchs,
Phys.~Rev.~Lett. \textbf{107}, 110501(2011).


\bibitem{WaveAtt}
H.~-W.~Li, S.~Wang, J.~-Z.~Huang, W.~Chen, Z.~-Q.~Yin, F.~-Y.~Li, Z.~Zhou, D.~Liu, Y.~Zhang, G.~-C.~Guo, W.~-S.~Bao, and Z.~-F.~Han,
Phys.~Rev.~A. \textbf{84}, 062308 (2011).


\bibitem{Weedbrook2011} C.~Weedbrook, S.~Pirandola, R.~Garc\'ia-Patr\'on, N.~J.~Cerf, T.~C.~Ralph, J.~H.~Shapiro, and S.~Lloyd,
Rev.~Mod.~Phys. \textbf{84}, 621 (2012).

\bibitem{Jouguet2012} P.~Jouguet, S.~Kunz-Jacques, A.~Leverrier, P.~Grangier, and E.~Diamanti, Nature Photonics. \textbf{7}, 378 (2013).

\bibitem{securenet} http://www.sequrenet.com/

\bibitem{quintessencelabs} http://qlabsusa.com/


\bibitem{cv}
F.~Grosshans, G.~van Assche, J.~Wenger, R.~Brouri, N.~J.~Cerf, and P.~Grangier, Nature. \textbf{421}, 238 (2003).

\bibitem{cv-het}
C.~Weedbrook, A.~M.~Lance, W.~P.~Bowen, T.~Symul, T.~C.~Ralph, and P.~K.~Lam, Phys.~Rev.~Lett. \textbf{93}, 170504 (2004);
C.~Weedbrook, A.~M.~Lance, W.~P.~Bowen, T.~Symul, T.~C.~Ralph, and P.~K.~Lam, Phys.~Rev.~A. \textbf{73}, 022316 (2006).

\bibitem{cv-exp}
J.~Lodewyck, T.~Debuisschert, R.~Tualle-Brouri, P.~Grangier, Phys.~Rev.~A. \textbf{72}, 050303 (2005).

\bibitem{testlo}
H.~H\"{a}seler, T.~Moroder and N.~L\"{u}tkenhaus, Phy.~Rev.~A. \textbf{77}, 032303 (2008).

\bibitem{cv-sec}
F.~Grosshans, N.~J.~Cerf, Phys.~Rev.~Lett. \textbf{92}, 047905 (2004).

\bibitem{het-sec}
J.~Lodewyck, and P.~Grangier,
Phys. Rev. A. \textbf{76}, 022332 (2007);
J.~Sudjana, L.~Magnin, R.~Garc\'{i}a-Patr\'{o}n, and N.~J.~Cerf,
Phys.~Rev.~A. \textbf{76}, 052301 (2007).

\bibitem{bs-0}
M.~Eisenmann, and E.~Weidel, J.~Lightwave~Tech. \textbf{6}, 8588-8594 (2010).

\bibitem{bs-1}
A.~Ankiewicz, A.~Snyder and X.~-H.~Zheng, J.~Lightwave~Tech. \textbf{4(9)}, pp1317-1323 (1986).

\bibitem{bs-2}
V.~Tekippe, Fiber~and~Integrated~Optics. \textbf{9(2)}, 97-123 (1990).

\bibitem{Grosshans2002} F.~Grosshans and P.~Grangier, Phys.~Rev.~Lett. \textbf{88}, 057902  (2002).

\bibitem{cv-rev}
S.~Braunstein and P.~van~Loock, Rev.~Mod.~Phys. \textbf{77}, 513 (2005).

\bibitem{book}
H.~Bachor and T.~C.~Ralph, ``A Guide to Experiments in Quantum Optics``, 2nd Edition, WILEY-VCH Press (2003).

\bibitem{cv-sec0}
F.~Grosshans, N.~J.~Cerf, J.~Wenger, R.~Tualle-Brouri, and P.~Grangier, Quantum~Inf.~Comput. \textbf{3}, 535 (2003).

\bibitem{GarciaPatron2007} R.~Garc\'{\i}a-Patr\'{o}n, Ph.D. thesis (Universit\'{e} Libre de Bruxelles) (2007).

\bibitem{note} In a regular CV QKD system, the LO intensity is approximately equal to the total intensity, therefore monitoring the total intensity is equivalent to monitor the LO intensity. See also ref.\cite{Jouguet2012}.

\bibitem{book2} G.~B.~Arfken and H.~J.~Weber, Mathematical Methods for Physics, Elesvier Academic Press, 2006.

%
%







\end{thebibliography}
\end{document}